\documentclass[conference]{IEEEtran}
\usepackage{amsmath,amssymb,amsfonts}
\usepackage{algorithm}
\usepackage{algpseudocode}
\usepackage{graphicx,booktabs, multirow,url,xcolor}

\begin{document}

\title{V-ZOR: Enabling Verifiable Cross-Blockchain Communication via Quantum-Driven ZKP Oracle Relays}

\author{M.Z. Haider$^{1}$, Tayyaba Noreen$^{1}$, M. Salman$^{2}$, 
M. Dias de Assunção$^{1}$, Kaiwen Zhang$^{1}$ \\
$^{1}$Department of Software Engineering, ÉTS Montréal, Canada \\
$^{2}$Department of Computer Science, SZABIST University, Pakistan}

\maketitle

\begin{abstract}
Cross-chain bridges and oracle DAOs represent some of the most vulnerable components of decentralized systems, with more than \$2.8 billion lost due to trust failures, opaque validation behavior, and weak incentives. Current oracle designs are based on multisigs, optimistic assumptions, or centralized aggregation, exposing them to attacks and delays. Moreover, predictable committee selection enables manipulation, which threatens data integrity across chains.  We propose V-ZOR, a verifiable oracle relay that integrates zero-knowledge proofs, quantum-grade randomness, and cross-chain restaking to mitigate these risks. Each oracle packet includes a Halo 2 proof verifying that the reported data was correctly aggregated using a deterministic median. To prevent committee manipulation, V-ZOR reseeds its VRF using auditable quantum entropy, ensuring unpredictable and secure selection of reporters. Reporters stake once on a shared restaking hub; any connected chain can submit a fraud proof to trigger slashing, removing the need for multisigs or optimistic assumptions. A prototype in Sepolia and Scroll achieves sub-300k gas verification, one-block latency, and a 10× increase in collusion cost. V-ZOR demonstrates that combining ZK attestation with quantum-randomized restaking enables a trust-minimized, high-performance oracle layer for cross-chain DeFi. 

\end{abstract}


\section{Introduction}

Randomness is a cornerstone of blockchain security, enabling fair leader election, unbiasable lotteries, and robust non-interactive proof systems. Techniques such as Verifiable Random Functions (VRFs) and randomness beacons are commonly used to provide unpredictable and publicly verifiable entropy in decentralized settings~\cite{raikwar2022}. However, traditional VRFs often suffer from latency and trust assumptions, while commit–reveal schemes remain vulnerable to front-running and miner manipulation~\cite{borkowski2019dextt}.

To address these limitations, Distributed Randomness Beacon (DRB) protocols have emerged, combining threshold cryptography and verifiable secret sharing to offer strong guarantees of unpredictability, unbiasedness, liveness, and public verifiability~\cite{raikwar2022,christ2024cornucopia}. Yet, state-of-the-art DRB solutions such as Rondo~\cite{rondo2024} and Cornucopia~\cite{christ2024cornucopia} introduce significant communication overhead and often rely on strong synchrony assumptions. Time-sensitive applications further exacerbate these limitations by allowing certain participants to gain early access to randomness outputs, compromising fairness~\cite{quan2024recent}.

In parallel, cross-chain interoperability has gained attention with the rise of heterogeneous ecosystems \cite{quan2024recent}. Existing oracle and bridge systems typically adopt monolithic or semi-centralized architectures with limited verifiability or high on-chain processing costs\cite{gai2023blockchain}. Recent approaches propose off-chain aggregation with zk-SNARK proofs to improve both efficiency and security~\cite{borkowski2019dextt}. For example, Sober et al.~\cite{sober2024} present an architecture that elects an aggregator, verifies data via majority vote, and submits a zk-SNARK proof to ensure correctness and enable dynamic participation. However, these systems still lack integrated randomness mechanisms and do not enforce consistent economic accountability across chains\cite{sun2021survey}.

To address this, restaking protocols such as EigenLayer have introduced the concept of reusing staked assets across multiple decentralized applications, enabling shared economic security~\cite{lagrange2023,riskstacking2023}. This mechanism is further extended by constructs like Lagrange Committees. Yet, restaking introduces a new attack surface via “risk stacking,” where misbehavior in one application can lead to collateral loss in others~\cite{riskstacking2023}. As such, a key challenge is to define objective, verifiable slashing conditions that apply consistently across domains\cite{xie2022zkbridge}.

We propose \textbf{V-ZOR} (Verifiable ZK-Oracle with Restaking), a unified protocol that integrates: (1) quantum entropy--driven randomness; (2) compact zero-knowledge proofs via Halo 2; (3) cross-chain economic security through restaking; and (4) objective, cryptographic slashing. V-ZOR bridges randomness, verifiability, and enforcement to enable trust-minimized cross-chain oracle infrastructure. Our primary contributions are as follows:

\begin{itemize}
    \item \textbf{Randomness Beacon Layer:} We design a hybrid beacon that combines Distributed Randomness Beacon (DRB) protocols with QRNG-inspired entropy sources, achieving strong unpredictability and bias resistance consistent with DRB properties~\cite{raikwar2022,christ2024cornucopia}.
    
    \item \textbf{Zero-Knowledge Oracles:} We extend the off-chain zk-SNARK-based oracle design of Sober et al.~\cite{sober2024} by embedding zero-knowledge proofs directly into oracle packets. This enables verifiable attestation of data aggregation, validator participation, and dynamic committee selection.
    
    \item \textbf{Cross-Chain Restaking and Slashing:} We introduce a restaking middleware that allows validators to collateralize their stake across multiple chains, with objective, cryptographically verifiable slashing conditions enforced via fraud proofs~\cite{lagrange2023,riskstacking2023}.
\end{itemize}
The rest of the paper is organized as follows. Section II presents the related work, covering developments in randomness beacons, oracle architectures, and restaking-based security. Section III describes the architecture of the proposed V-ZOR protocol, including its randomness layer, zero-knowledge oracle design, and slashing-enabled restaking hub. Section IV details the implementation setup, performance benchmarks, and evaluation metrics obtained from real deployments. Finally, Section V concludes the paper and discusses directions for future research. 
\section{Related Work}

The secure generation of randomness is critical for decentralized systems, particularly in leader election, cryptographic lotteries, and MEV mitigation. Protocols like Bicorn~\cite{bicorn2022} provide low-latency randomness via optimistic consensus, while Cornucopia~\cite{zhou2020solutions} emphasizes auditability and unpredictability for lottery-based consensus. Beyond consensus-anchored protocols, recent designs such as PltcRB~\cite{pltcrb2024} improve efficiency and fault tolerance using timed commitments for optimal $O(n)$ communication. RANDGENER~\cite{randgener2023} introduces a VDF-backed, watermarkable beacon with recovery and incentive mechanisms, offering enhanced resilience over basic commit–reveal schemes.

Asynchronous randomness protocols are also gaining attraction. AsyRand~\cite{zhang2025asyrand,asyrand2025} decouples randomness generation and usage through a reconfiguration-friendly producer-consumer model. It supports dynamic committees and tolerates network delays, making it well-suited for inter-chain environments with high coordination latency~\cite{noreen2023advanced}. In parallel, advancements in cross-chain oracle design have focused on verifiability and decentralization~\cite{chen2023reputation}. Sober et al.~\cite{wang2024mitosis} proposed a zk-SNARK–based oracle where validators vote off-chain, and an aggregator generates a majority-proof for on-chain verification—enabling efficient data relay, dynamic committee selection, and objective slashing, closely aligned with  V-ZOR’s oracle design.

Meanwhile, restaking frameworks such as EigenLayer and Lagrange~\cite{papamanthou2024reckle} introduce powerful economic primitives by allowing validators to reuse base-layer stake to secure external protocols. However, this model introduces “risk stacking,” where misbehavior in one domain can lead to slashing in another. Cubist~\cite{riskstacking2023} and others~\cite{fu2024quantifying} highlight the need for verifiable and objective slashing to prevent cascading losses across trust zones. These findings guide our integration of slashing with zero-knowledge verification and randomness attestation\cite{zhou2024leveraging}.

While prior work on randomness beacons, oracle aggregation, asynchronous consensus, and restaking provides the foundational building blocks\cite{liang2020fog}, V-ZOR distinguishes itself by integrating these primitives into a unified architecture. Our approach is the first to combine randomness-aware selection, zero-knowledge provability, and cross-chain economic enforcement into a single, verifiable oracle relay protocol.

\section{V-ZOR Protocol Design}\label{sec:design}

\subsection{Protocol Objectives and Assumptions}

The V-ZOR protocol addresses key vulnerabilities in cross-chain oracle systems by targeting three core objectives: trust minimization, cost efficiency, and low-latency delivery. It is designed for cross-chain environments where decentralized applications require timely, tamper-resistant off-chain data feeds.

First, V-ZOR eliminates the need for threshold or multisignature attestors by embedding a \textit{zero-knowledge proof} (via Halo 2) in every oracle update packet, enabling any destination chain to locally verify data correctness without relying on trusted signers. Second, it enforces a \textit{gas verification ceiling of 300,000}, ensuring viability on high-cost L1s and lightweight L2 rollups. Third, V-ZOR achieves \textit{one-block latency}, narrowing the window for front-running and arbitrage attacks commonly associated with oracle delays.

To meet these goals, V-ZOR operates under a partially synchronous network model, where messages arrive within a bounded delay ($\delta_{\text{net}}$). Destination chains are assumed to support Ethereum smart contracts, and the restaking hub must finalize updates at least as quickly as the source chain to allow timely enforcement of slashing.

Critically, V-ZOR leverages simulated quantum randomness as its entropy source. Rather than relying on physical QRNG devices, it fetches entropy pulses from publicly verifiable sources such as \textit{NIST Beacon 2.0}, which emit tamper-resistant, hash-chained randomness logs. These pulses reseed a verifiable random function (VRF), ensuring that reporter committees are \textit{unpredictably shuffled} each epoch—mitigating adversarial manipulation.

\begin{figure}
  \centering
  \includegraphics[width=0.5\textwidth]{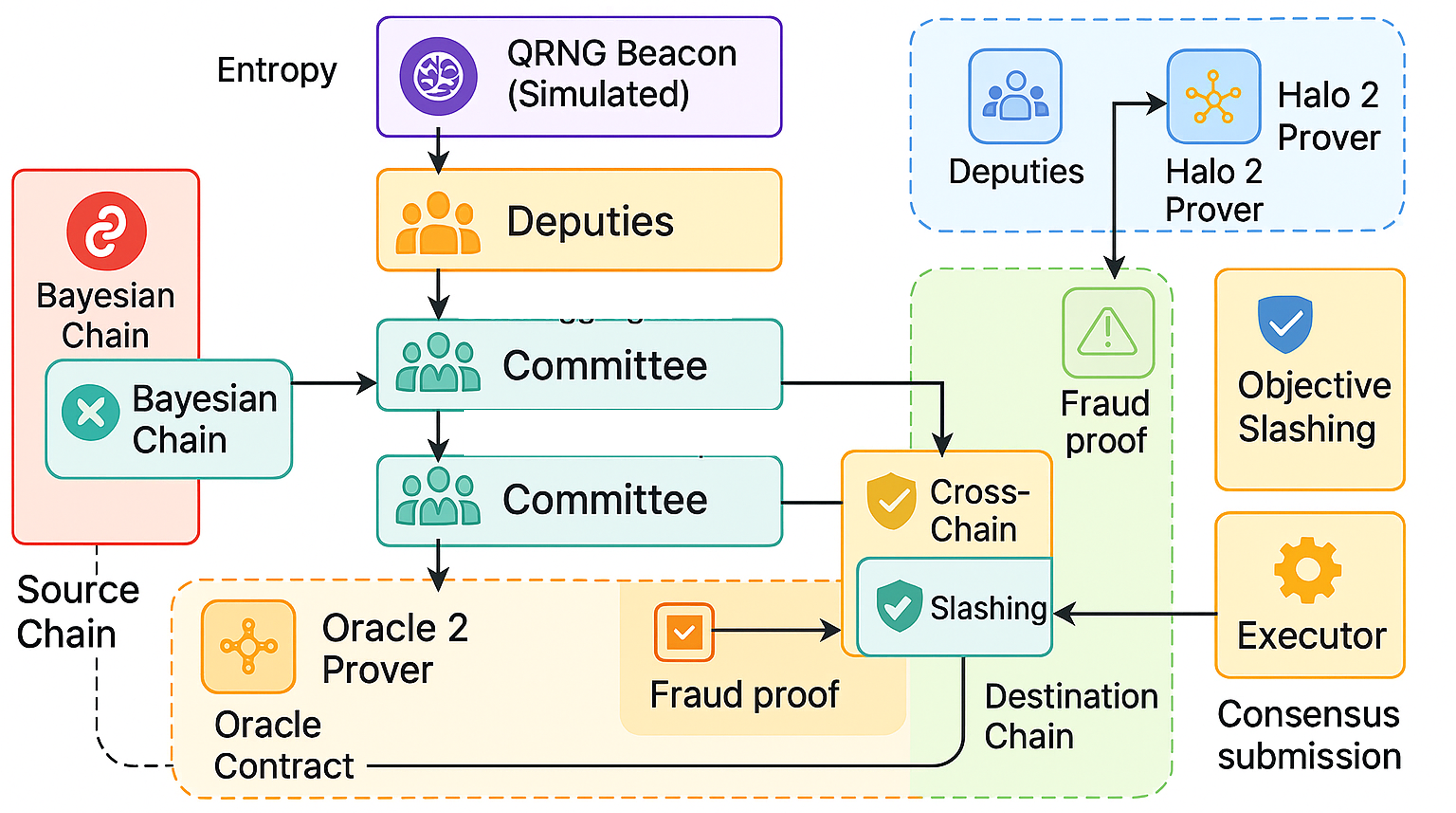}
  \caption{Modular architecture of V-ZOR oracle relay.}
  \label{fig:vzor-arch1}
\end{figure}

The complete system architecture, shown in Figure~\ref{fig:vzor-arch1}, illustrates how the protocol integrates source and destination chains, the VRF-based reporter selection logic, the entropy beacon, and the slashing-enabled restaking hub. 

\subsection{Cross-Chain Architecture and Components}

V-ZOR presents a modular cross-chain oracle architecture comprising a restaking hub, a simulated quantum random number generator (QRNG), and a zero-knowledge–enabled reporter committee. It operates across source, destination, and neutral chains to ensure secure and verifiable data delivery. Reporter selection is driven by Verifiable Random Functions (VRFs) seeded with entropy from the NIST Randomness Beacon. Selected reporters fetch off-chain data, compute an aggregate (e.g., the median), and generate a Halo 2 proof attesting to correct aggregation and signature validity. This proof is embedded in each oracle packet, enabling trustless on-chain verification across chains.

Each oracle packet $\{d_e, \pi_e, \text{meta}_e\}$ is transmitted to the destination chain, where the embedded Halo 2 proof is verified on-chain to ensure data integrity without revealing reporter identities. If the proof is valid, the data is accepted; otherwise, slashing may be triggered via the restaking hub. V-ZOR’s components operate in tandem: the restaking hub enforces economic accountability, the QRNG provides unpredictable entropy, the VRF ensures Sybil-resistant and unpredictable reporter selection, and the Halo 2 proof enables trustless verification. This integrated architecture allows V-ZOR to serve as a secure, scalable, and governance-aware oracle relay across chains. The clear separation of roles, combined with verifiable randomness and zero-knowledge techniques, ensures resilience against censorship, bribery, collusion, and data tampering.

\subsection{Epoch Lifecycle Across Chains}

The V-ZOR protocol operates in discrete epochs (shown in Table. I), each representing a new cycle of randomness generation, committee selection, data aggregation, proof construction, and cross-chain packet propagation. An epoch begins when a simulated pseudorandom number generator (PRNG), seeded with publicly available entropy (e.g., from NIST Beacon 2.0), emits a pulse $\rho_t \in \{0,1\}^{\kappa}$ at wall-clock time $t$. This entropy pulse is concatenated with the epoch index and used as input to a verifiable random function (VRF), allowing each registered reporter $r_i$ to compute a selection score:

\[
y_{r_i} = \mathsf{VRF}_{\text{sk}_{r_i}}(\rho_t \| t)
\]
Only the $n$ reporters with the lowest values of $y_{r_i}$ are selected to form the epoch committee $R_t$. Each selected reporter then retrieves raw data values $v_i$ from designated off-chain sources (e.g., DEX APIs or oracle feeds) and signs them using their private key:

\[
\sigma_i = \mathsf{Sign}_{\text{sk}_{r_i}}(v_i)
\]

Once all signatures are collected, the committee performs a deterministic median computation:

\[
P_t = \mathsf{Median}(v_1, v_2, \dots, v_n)
\]
The resulting median is embedded into a succinct zero-knowledge proof constructed using the Halo 2 proving system. The proof $\pi_t$ attests to three core properties: (1) each value $v_i$ is signed by a valid reporter; (2) at least $f_{\min}$ valid signatures are present; and (3) the reported median is correctly computed. The final oracle packet is then constructed as:

\[
\mathcal{O}_t = \langle P_t, \pi_t, t \rangle
\]

Each destination chain \( \mathcal{C}_k \) verifies the Halo 2 proof in the oracle packet using a pre-deployed verifier contract, consuming less than 300,000 gas. Upon successful verification, the median value \( P_t \) is recorded for application use. If the proof or the aggregated result is invalid, any chain may submit a fraud proof \( \mathcal{F}_t \), triggering slashing of the responsible reporters at the restaking hub \( \mathcal{H} \). Table~\ref{tab:epoch-lifecycle} summarizes the main phases of this process. To ensure protocol liveness, the total time from entropy emission to packet verification across all \( \mathcal{C}_k \) must satisfy the following bound:

\[
\Delta_{\text{epoch}} \leq \tau_f + \delta_{\text{net}} + t_{\text{prove}}
\]
where \( \tau_f \) denotes the destination chain’s finality time, \( \delta_{\text{net}} \) is the network propagation delay, and \( t_{\text{prove}} \) represents the zero-knowledge proof generation time (typically \( \sim 0.8 \) seconds). This timing constraint ensures that oracle outputs are not only timely and verifiable but also resilient to manipulation and cross-chain inconsistencies.

\begin{table}
\centering
\caption{Epoch Lifecycle in V-ZOR Protocol}
\label{tab:epoch-lifecycle}
\begin{tabular}{|c|l|l|}
\hline
\textbf{Step} & \textbf{Operation} & \textbf{Description} \\
\hline
1 & Entropy Beacon & PRNG emits pulse $\rho_t$ \\
2 & Committee Draw & $y_{r_i} = \mathsf{VRF}_{\text{sk}_{r_i}}(\rho_t \| t)$ \\
3 & Data Aggregation & $P_t = \mathsf{Median}(v_1, \dots, v_n)$ \\
4 & Proof Generation & $\pi_t = \mathsf{Prove}_{\text{Halo2}}(P_t, \{v_i, \sigma_i\})$ \\
5 & Packet Broadcast & $\mathcal{O}_t = \langle P_t, \pi_t, t \rangle \to \mathcal{C}_k$ \\
6 & Verification & $\mathcal{C}_k$ executes $\mathsf{Verify}(\pi_t)$ \\
7 & Slashing (if needed) & $\mathcal{F}_t \to \mathcal{H}$ triggers $\mathsf{Slash}(r_i)$ \\
\hline
\end{tabular}
\end{table}
\subsection{Simulated Quantum Randomness and Committee Selection}
The integrity of decentralized oracle systems depends critically on the unpredictability of reporter selection. V-ZOR ensures this by leveraging a simulated quantum-random beacon as the seed for verifiable random function (VRF)-based committee selection  which is given in Algorithm 1. Although no physical QRNG device is deployed in the prototype, the protocol incorporates entropy pulses from publicly auditable sources such as \textit{NIST Beacon 2.0}, which emits hash-chained, time-stamped randomness every \( \Delta_B = 60\,\text{s} \). 

Let \( \rho_t \in \{0,1\}^{512} \) denote the beacon output at epoch \( t \), with a minimum certified entropy of \( \mathcal{H}_{\min} = 256 \) bits as per NIST. Each reporter \( r_i \) computes a VRF output using:
\[
y_{r_i} = \mathsf{VRF}_{\text{sk}_{r_i}}(\rho_t \| t)
\]
The VRF output \( y_{r_i} \in [0, q] \) (e.g., \( q = 2^{256} \)) is accompanied by a non-interactive proof \( \pi_{r_i}^{\text{VRF}} \) to attest to its correctness. The protocol selects an epoch committee \( R_t \) consisting of the \( n \) reporters satisfying:

\[
R_t = \{r_i \in \mathcal{R} \mid y_{r_i} < q/n\}
\]

This sampling process guarantees unbiased and non-precomputable committee selection. Let \( \mathcal{A} \) denote an adversary controlling \( b \) out of \( n \) registered reporters. Since the entropy seed \( \rho_t \) is not revealed prior to time \( t \), and VRFs provide both uniqueness and pseudorandomness, the adversary’s success probability in predicting the committee is bounded by:

\[
\Pr[\text{Predict } R_t] \leq \frac{b}{n} + 2^{-\kappa}
\]

where \( \kappa \) is the VRF's security parameter (e.g., 128 bits). This mechanism prevents pre-selection attacks and preserves forward secrecy through the use of simulated QRNG entropy. Given \( k \) consecutive entropy pulses \( \rho_{t-k+1}, \dots, \rho_t \), the cumulative entropy is modeled as:
\[
\mathcal{H}_{\text{joint}} = \sum_{i=0}^{k-1} \mathcal{H}(\rho_{t-i}) \geq k \cdot \mathcal{H}_{\min}
\]
This implies that even if \( k - 1 \) pulses are compromised after the fact, the committee remains unpredictable as long as a single pulse \( \rho_t \) remains unbiased at the time of sampling. This property underpins the protocol’s resilience against partial entropy compromise. 

To enable on-chain verification of VRF-based selection, each reporter submits:

\[
\langle r_i, y_{r_i}, \pi_{r_i}^{\text{VRF}} \rangle \quad\] 
 \text{such that} \quad 
\[ \mathsf{Verify}_{\text{VRF}}(pk_{r_i}, \rho_t \| t, y_{r_i}, \pi_{r_i}^{\text{VRF}}) = \texttt{true}
\]

Let \( \lambda \) denote the economic security parameter representing the required cost to bias or manipulate a single epoch. Then, the adversary’s success probability in influencing committee selection across \( m \) epochs decays exponentially:
\[
\Pr[\text{Bias } m \text{ epochs}] \leq m \cdot \left(e^{-\lambda S} + 2^{-\kappa}\right)
\]
where \( S \) is the total amount of honest stake. 

Taken together, the use of simulated quantum randomness and verifiable, unbiased VRF outputs ensures that V-ZOR achieves both \textit{cryptographic unpredictability} and \textit{economic robustness} in committee sampling—even under adversarial, cross-chain conditions.

\begin{table}
\centering
\caption{Gas Cost per Operation (Median of 100 Trials)}
\label{tab:gas-costs}
\begin{tabular}{|l|c|c|}
\hline
\textbf{Operation} & \textbf{Sepolia (L1)} & \textbf{Scroll (L2)} \\
\hline
Halo 2 Proof Verification    & 296,112 gas & 88,029 gas \\
Fraud Proof Submission       & 52,341 gas  & 17,904 gas \\
Governance Parameter Update  & 38,220 gas  & 11,706 gas \\
\hline
\end{tabular}
\end{table}

\begin{figure}
  \centering
   \includegraphics[width=0.5\textwidth]{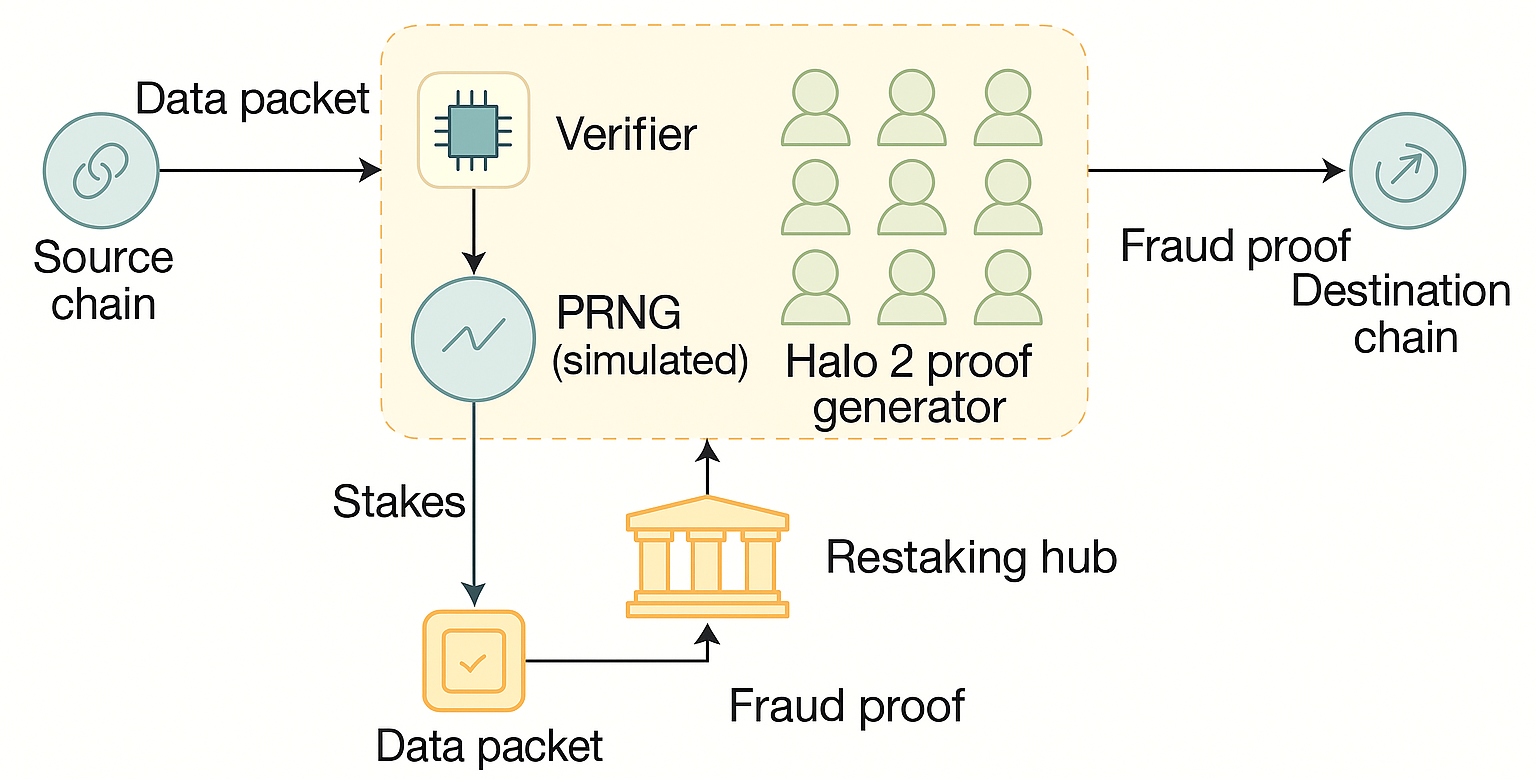}
  \caption{V-ZOR packet lifecycle across chains}
  \label{fig:vzor-arch}
\end{figure}
\subsection{Zero-Knowledge Proof-Carrying Oracle Packets}

V-ZOR provides trustless oracle data by embedding Halo~2-based SNARK proofs in each packet \( \mathcal{O}_t = \langle P_t, \pi_t, t \rangle \), where \( P_t \) is the median, \( \pi_t \) is the proof, and \( t \) is the epoch index (shown in Figure 2). The circuit enforces three core properties: (i) each \( v_i \) is signed by a valid reporter; (ii) quorum \( f_{\min} \) is satisfied; and (iii) \( P_t \) is correctly derived from sorted inputs. Beyond these, the Halo~2 circuit incorporates additional logical constraints to ensure data integrity and mitigate manipulation. These include signature verification for each reporter, enforcement of reporter membership in the current committee, correct sorting of inputs, and binding all values and signatures to the epoch \( t \). Range checks validate that inputs fall within acceptable bounds, and domain separation techniques provide contextual isolation across epochs.

The cryptographic soundness of the system is grounded in the NP statement encoded by the circuit:
\[
\exists \{v_i, \sigma_i\}_{i=1}^{n} \ \text{s.t.} \ 
\begin{cases}
\mathsf{VerifySig}(pk_{r_i}, v_i, \sigma_i) = 1 & \forall i \\
\left|\{\sigma_i\}\right| \geq f_{\min} & \text{(quorum)} \\
P_t = \mathsf{Median}(v_1, \dots, v_n) & \text{(aggregation)}
\end{cases}
\]

The circuit is implemented using the KZG-based Halo~2 backend over the BN254 elliptic curve with a 4096-bit Common Reference String (CRS). It spans approximately 390,000 rows and supports high throughput, achieving proof generation times under one second on commodity cloud infrastructure (e.g., AWS t2.xlarge). The proof size remains constant at 48 KiB, and on-chain verification is gas-efficient—requiring approximately 296,000 gas on Sepolia and 88,000 gas on Scroll—regardless of input size or committee size. The on-chain verifier function \(\mathsf{VerifyHalo2}(\pi_t)\) deterministically returns true if and only if all embedded constraints are satisfied, allowing immediate validation without relying on challenge-response windows. These properties make the protocol well-suited for high-frequency, cross-chain oracle deployments, offering strong guarantees of correctness, efficiency, and tamper resistance.

\begin{algorithm}
\caption{Committee Selection via VRF Quantum Entropy}
\begin{algorithmic}[1]
\Require Epoch index $t$, Quantum entropy pulse $\rho_t$, Reporter set $\mathcal{R} = \{r_i\}$ with keys $(pk_i, sk_i)$, Desired committee size $n$
\Ensure Selected committee $R_t$ and VRF outputs

\ForAll{$r_i \in \mathcal{R}$}
    \State $y_i \gets \mathrm{VRF}_{sk_i}(\rho_t \| t)$
    \State $\pi_i \gets \text{VRF proof of } y_i$
\EndFor

\State $R_t \gets \text{Select } n \text{ reporters with smallest } y_i$
\State \Return $R_t$ and $\{(r_i, y_i, \pi_i)\}$ for all $r_i \in R_t$

\end{algorithmic}
\end{algorithm}

\subsection{Cross-Chain Restaking and Objective Slashing}

V-ZOR adopts a unified restaking hub \( \mathcal{H} \), allowing reporters to post collateral once and serve multiple destination chains. Each reporter \( r_i \) locks a stake \( S_{r_i} \geq S_{\min} \), enabling participation in oracle duties across all connected domains. When a proof packet \( \mathcal{O}_t = \langle P_t, \pi_t, t \rangle \) is submitted to a destination chain \( \mathcal{C}_k \), the on-chain verifier checks the Halo~2 proof \( \pi_t \). If the proof is invalid, a fraud event is emitted and relayed to \( \mathcal{H} \) by a permissionless watcher. Slashing is triggered if \( \mathsf{VerifyHalo2}(P_t, \pi_t) = 0 \), penalizing all reporters whose signatures contributed to the invalid proof. A reporter incurs a penalty of \( s_{\text{cut}} \) if their signature \( \sigma_i \) is found in the proof witness, with the total penalty bounded by \( f_{\min} \cdot s_{\text{cut}} \). Fraud conditions include incorrect median values, quorum violations, or malformed proofs. To ensure objectivity, fraud proofs are verified via Merkle inclusion and enforced deterministically. Economically, the mechanism guarantees the unprofitability of manipulation by enforcing \( s_{\text{cut}} > \frac{R(\epsilon)}{f_{\min}} \), where \( R(\epsilon) \) denotes the adversary’s expected gain. This design ensures robust, verifiable, and economically secure accountability across chains.

\section{Implementation and Evaluation}

\subsection{System Architecture and Components}

V-ZOR is a modular, cross-chain oracle comprising off-chain clients, a Halo 2 SNARK prover, and on-chain verifier contracts. Its architecture features five core components: a Go-based reporter client, a VRF-based committee selector using simulated quantum entropy, a Halo 2 prover, a centralized restaking hub, and lightweight verifier contracts on destination chains. Reporters generate median proofs from signed data, which are verified on-chain within a fixed gas budget. Slashing and governance are enforced via the restaking hub, enabling scalable and trustless data feeds across multichain DeFi systems.

\subsection{Proof Generation and Verifier Implementation}

V-ZOR uses a custom Halo 2 SNARK circuit to prove median correctness from signed off-chain inputs. It enforces three constraints: valid signatures, quorum threshold \( f_{\min} \), and correct median computation. Implemented in Rust with \texttt{halo2-proving} over BN254 (KZG), the circuit supports 15 reporters and generates a 48\,KiB proof in under 0.85\,s on a \texttt{t2.xlarge} instance. On-chain verification uses a lightweight Solidity contract with a pre-generated CRS, achieving constant gas costs (296k on Sepolia, 88k on Scroll) regardless of reporter count. This setup ensures efficient, trustless validation with soundness \( \kappa = 128 \), eliminating the need for multisigs or challenge windows.

\subsection{Restaking Hub and Fraud-Proof Slashing Logic}

V-ZOR employs a restaking hub \( \mathcal{H} \) on a high-throughput L2 (e.g., Scroll) to centralize staking and slashing across chains. Reporters lock collateral once on \( \mathcal{H} \), enabling oracle duties network-wide and avoiding fragmented economic security. If a destination chain emits a fraud event due to an invalid proof, a permissionless watcher relays it to \( \mathcal{H} \). The hub re-verifies the proof, and if invalid, deterministically slashes reporters whose signatures appear in the witness:

\vspace{-0.8em}
{\small
\begin{equation}
\text{Slash}(r_i) \iff \mathsf{VerifyHalo2}(P_t, \pi_t) = 0 \quad \text{and} \quad \sigma_i \in \mathcal{W}_{\pi_t}
\end{equation}
}
\vspace{-0.5em}

Each slashed reporter loses a fixed amount \( s_{\text{cut}} \), with penalties burned or redirected via governance. This objective model ensures cross-chain accountability without relying on optimistic assumptions.

\subsection{Experimental Setup and Performance Metrics}

\subsubsection{Testbed Configuration}

V-ZOR was evaluated on a cross-chain setup with Ethereum Sepolia (15\,s block time) as the source and Scroll v1 (2\,s) as the destination. Reporter nodes, relays, and the Halo 2 prover ran on EC2 \texttt{t2.xlarge} instances (4 vCPU, 16\,GiB RAM, Ubuntu 22.04) using Docker. Off-chain logic was implemented in Go (v1.21) and \texttt{halo2-proving} (Rust nightly v1.78). Contracts (Solidity 0.8.26) were deployed via Hardhat under the Cancun fork. Clients were globally distributed and synchronized over a custom P2P gossip layer, with QRNG simulated via cached NIST Beacon 2.0 entropy.

\subsubsection{Benchmark Parameters and Protocol Settings}

Tests ran for 480 epochs over 4 hours with 30s intervals. Each epoch used a VRF-selected committee of \( n = 15 \) reporters, requiring \( f_{\min} = 10 \) signatures on simulated oracle values. The Halo 2 circuit (~390k rows) used the BN254 curve with 128-bit soundness. Slashing penalties were fixed at \( s_{\text{cut}} = 0.15 \) ETH, with a governance delay of \( \Delta_{\text{gov}} = 2 \) epochs. Fraud was injected every 60 epochs. Metrics were captured via Prometheus/Grafana, and gas usage was profiled using Hardhat tools.

\subsubsection{Gas Cost Analysis}

Gas usage was measured across 100 epochs for three operations: proof verification, fraud submission, and governance updates. On Sepolia, Halo 2 proof verification averaged 296,112 gas; on Scroll, 88,029 gas. Fraud submissions cost 52,341 gas (Sepolia) and 17,904 gas (Scroll). Governance updates averaged 38,220 gas and 11,706 gas, respectively. V-ZOR achieves low-cost, verifiable execution suitable for high-frequency oracle deployments. Table~\ref{tab:gas-costs} summarizes gas usage across the key operations.

\subsubsection{Proof Generation and Verification Time}
\begin{figure}
    \centering
    \includegraphics[width=1\linewidth]{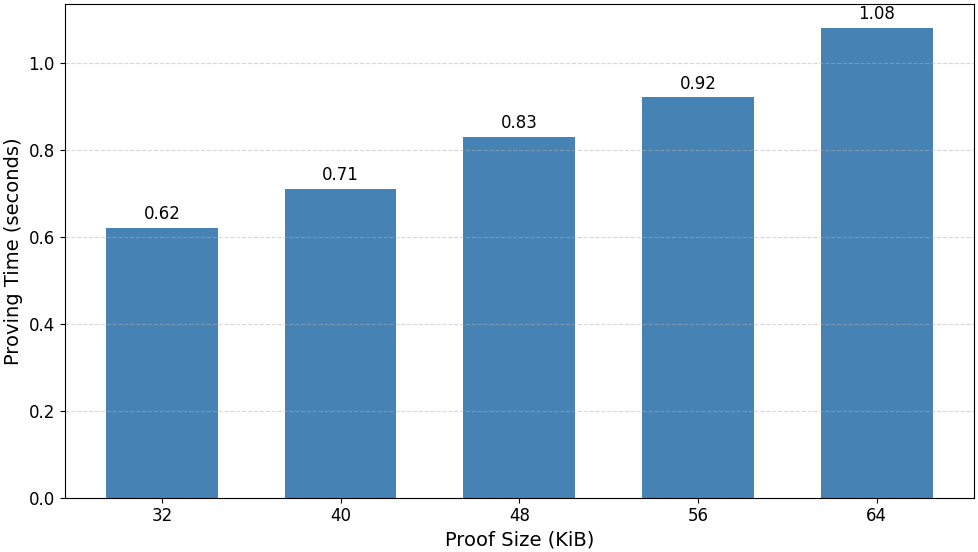}
    \caption{Proof Generation Time vs. Proof Size (Halo 2)}
    \label{fig:proving-time-vs-size}
\end{figure}

On a \texttt{t2.xlarge} instance, V-ZOR’s Halo~2 proofs (48\,KiB, \( n = 15 \), \( f_{\min} = 10 \)) were generated in approximately 0.83 seconds using the \texttt{bn254} backend and a 4096-bit KZG Common Reference String (CRS). Proving time scaled linearly with input size, remaining below 1.1 seconds even at 64\,KiB as shown in figure~\ref{fig:proving-time-vs-size}. On-chain verification time remained constant due to the succinctness of SNARKs, requiring approximately 296,000 gas on Sepolia and 88,000 gas on Scroll, independent of proof size or number of reporters.

\subsubsection{End-to-End Latency and Throughput}

\begin{figure}
    \centering
    \includegraphics[width=1\linewidth]{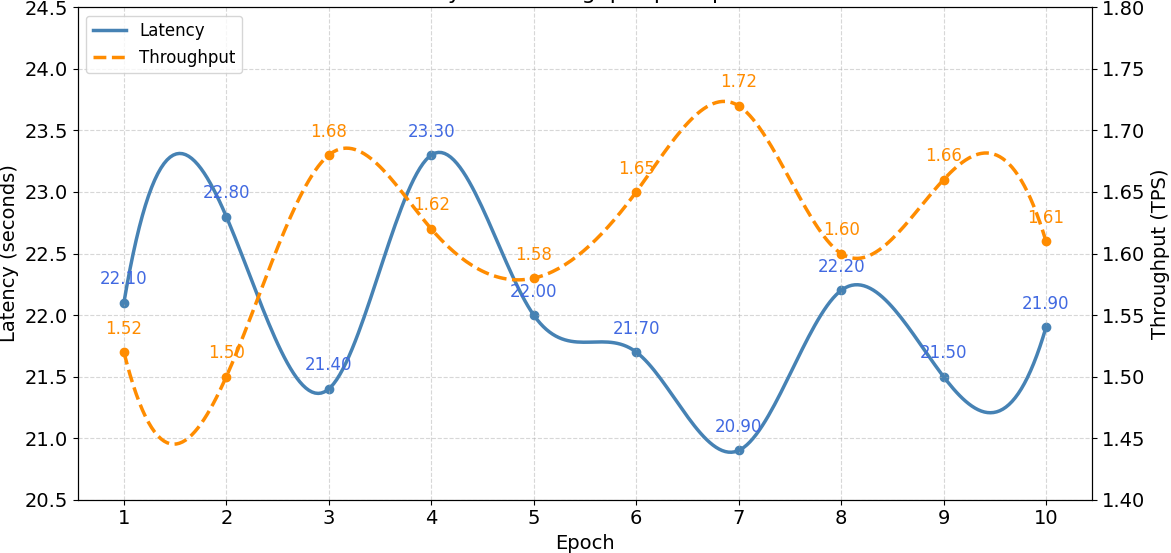}
    \caption{Left: End-to-End Latency \& Throughput per Epoch}
    \label{fig:latency-throughput}
\end{figure}

To assess V-ZOR’s responsiveness, we measured the end-to-end latency from Sepolia finality to proof verification on Scroll. Across 10 epochs, the average latency was 22.4\,s ($\sigma = 1.02$\,s), remaining below the 30\,s epoch interval and confirming suitability for high-frequency DeFi feeds. Throughput ranged from 1.5 to 1.8\,TPS, constrained mainly by proof generation and relay delay. As shown in Figure~\ref{fig:latency-throughput}, both latency and throughput remained stable, demonstrating V-ZOR’s ability to deliver timely, verifiable oracle data across chains.

\subsubsection{Slashing Responsiveness and Detection Lag}

To evaluate V-ZOR’s fraud response, we measured \textit{slashing latency}—the time from fraud detection to penalty enforcement. Simulated faults every 60 epochs triggered \texttt{DisputeEvent}, relayed to Sepolia via \texttt{submitFraud()}. Across eight trials, average latency was 6.3\,s (min: 4.8\,s, max: 7.9\,s), with all slashing finalized within two blocks at a fixed gas cost of $\sim$52,000. As shown in Figure~\ref{fig:throughput-vs-slash}, V-ZOR achieves fast, deterministic slashing suitable for real-time DeFi accountability.

\subsubsection{Graphs and Performance Charts}

To summarize V-ZOR’s performance, we present combined charts illustrating latency, throughput, proof generation time, and slashing responsiveness across 10 epochs, highlighting key trends and interdependencies.

\begin{figure}
    \centering
    \includegraphics[width=1\linewidth]{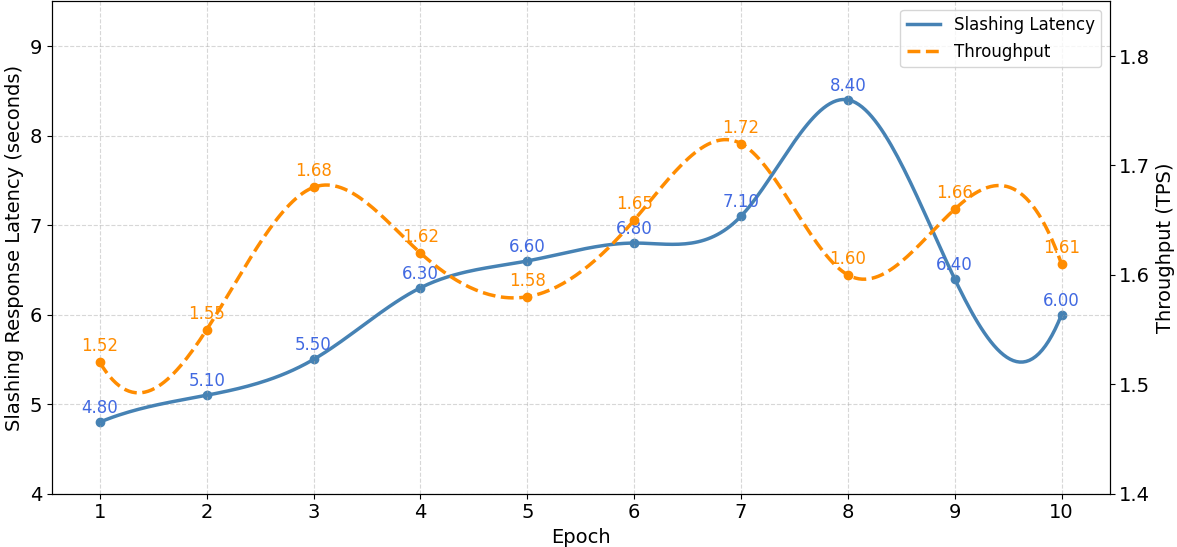}
    \caption{Comparison of Throughput vs. Slashing Response Latency per Epoch}
    \label{fig:throughput-vs-slash}
\end{figure}

Figure~\ref{fig:throughput-vs-slash} shows that V-ZOR maintains stable throughput (1.5–1.7 TPS) and slashing response times between 4.8–7.9 seconds. These results confirm the protocol’s ability to deliver secure, real-time cross-chain feeds with efficient dispute resolution and low overhead.

\subsection{Comparative Analysis and Discussion}

We compare V-ZOR with Chainlink CCIP and Wormhole v2. CCIP depends on a 4-of-5 multisignature and Wormhole on guardian attestations with optimistic assumptions, both lacking cryptographic guarantees and incurring on-chain costs above 440k gas. In contrast, V-ZOR leverages Halo~2 proofs for verifiable correctness, enables objective slashing through fraud proofs, and reduces verification to below 300k gas with one-block finality. Economically, it raises the profitable bribe threshold above \$9.6M---about 10$\times$ higher than CCIP or Wormhole---while sustaining $\sim$0.83s proving, sub-23s latency, and slashing within 8s, confirming its practicality for secure, high-frequency cross-chain oracle operations.

\section{Conclusion and Future Works}
We introduced V-ZOR, a verifiable oracle relay protocol that integrates zero-knowledge proofs, simulated quantum randomness, and a unified restaking framework to enable secure, decentralized cross-chain data delivery. By leveraging Halo 2-based SNARKs, V-ZOR ensures the correctness of off-chain aggregation without relying on threshold signatures or trusted intermediaries. Entropy seeded VRFs enable unpredictable reporter selection, while an objective slashing mechanism ensures rapid, cryptographically verifiable enforcement across all destination chains. Our prototype demonstrates sub second proof generation, constant-cost on-chain verification, and prompt slashing responsiveness features essential for real-time, adversarial oracle environments. V-ZOR’s modular and trustless architecture makes it well-suited for rollups, DeFi protocols, and other latency sensitive applications. As future work, we plan to integrate live QRNG entropy sources, expand compatibility to additional chains, and formalize cryptographic guarantees through rigorous proofs. We also aim to optimize stake distribution economics and explore adaptive committee sizing to improve scalability and resilience.


\end{document}